\documentclass[pra,preprint,showpacs,showkeys,nofootinbib,tightenlines]{revtex4}

\usepackage{graphicx}  %
\usepackage{bm}  %

\newcommand{\simge}{\hspace*{0.2em}\raisebox{0.5ex}{$>$}
     \hspace{-0.8em}\raisebox{-0.3em}{$\sim$}\hspace*{0.2em}}

\def\vec#1{{\bf #1}}

\newcommand{\beq}{\begin{equation}}
\newcommand{\eeq}{\end{equation}}
\newcommand{\bqa}{\begin{eqnarray}}
\newcommand{\eqa}{\end{eqnarray}}

\def\mqo2{{\!\!\!}}

\begin{document}

\title{Universal Equation for Efimov States}

\author{Eric Braaten}\email{braaten@mps.ohio-state.edu}
\author{H.-W. Hammer}\email{hammer@itkp.uni-bonn.de}\thanks{
 Present address: Helmholtz-Institut f{\"u}r Strahlen- und Kernphysik
 (Abt. Theorie), Universit{\"a}t Bonn, 53115 Bonn, Germany}
\author{M. Kusunoki}\email{masa@mps.ohio-state.edu}

\affiliation{Department of Physics,
         The Ohio State University, Columbus, OH\ 43210, USA}

\date{November 22, 2002}

\begin{abstract}
Efimov states are a sequence of shallow 3-body
bound states that arise when the 2-body scattering length is large. 
Efimov showed that the binding energies of these states 
can be calculated in terms of the scattering length and a 3-body 
parameter by solving a transcendental equation involving
a universal function of one variable. 
We calculate this universal function using effective field theory
and use it to describe the three-body system of $^4$He atoms.
We also extend Efimov's theory to include the effects of deep 2-body
bound states, which give widths to the Efimov states.
\end{abstract}

\smallskip
\maketitle  

The interactions of nonrelativistic particles (such
as atoms) with short-range interactions at extremely low energies 
are determined primarily by their $S$-wave scattering length $a$. 
If $|a|$ is much larger 
than the characteristic range $l$ of their interaction, low-energy
atoms exhibit universal properties that are insensitive to the 
details of the interaction potential. In the 2-body sector, the
universal properties are simple and familiar. The differential cross 
section for two identical bosons with relative wave number $k \ll 1/l$ 
and mass $m$ is $d\sigma/d\Omega=4a^2/(1+k^2 a^2)$. If $a>0$, there
is also a shallow 2-body bound state
(the dimer) with binding energy $B_2=\hbar^2/ma^2$.
In the 3-body sector, there are also universal properties that
were first deduced by Efimov \cite{Efi71}.
The most remarkable is the existence of a sequence of 3-body 
bound states with binding energies geometrically spaced in the interval 
between $\hbar^2/ma^2$ and $\hbar^2/ml^2$. The number of these
\lq\lq Efimov states'' is roughly $\ln(|a|/l)/\pi$
if $|a|$ is large enough. In the limit
$|a|\to \infty$, there is an accumulation of infinitely many
3-body bound states at threshold (the \lq\lq Efimov effect'').
The knowledge of the Efimov binding energies is
essential for understanding the energy-dependence of low-energy 3-body
observables. For example, Efimov states can have dramatic 
effects on atom-dimer scattering if $a>0$ \cite{Efi71,BHK99}
and on 3-body recombination if $a<0$ \cite{EGB99,BBH00}. 

A large 2-body scattering length can be obtained by fine-tuning a
parameter in the interatomic potential to bring a 
real or virtual 2-body bound 
state close to the 2-atom threshold. The fine tuning can
be provided accidentally by nature. An example is the $^4$He atom,
whose scattering length $a= 104$ \AA\ \cite{Gri00}
is much larger than the effective range $l\approx 7$ \AA.
Another example is the 2-nucleon system in the $^3 S_1$
channel, for which the deuteron is the shallow bound state.
This system provided the original motivation for
Efimov's investigations \cite{Efi71}. In the case of atoms,
the fine tuning can also be obtained by tuning an external electric 
field \cite{NFJ99} or by tuning an external
magnetic field to the neighbourhood of a Feshbach resonance
\cite{Festh}. Such resonances have, e.g., been observed for
$^{23}$Na and $^{85}$Rb atoms \cite{Fesex} and are used
to tune the interactions in Bose-Einstein condensates.
An important difference from He is that the interatomic 
potentials for Na and Rb support many deep 2-body bound states.

Efimov derived some powerful constraints on low-energy 3-body
observables for systems with large scattering length \cite{Efi71}.
They follow from the approximate scale-invariance at 
length scales $R$ in the region $l \ll R \ll |a|$ and the 
conservation of probability. 
He introduced polar variables $H$ and $\xi$ in
the plane whose axes are $1/a$ and the energy variable
${\rm sgn}(E)|mE|^{1/2}/\hbar$, and showed that 
low-energy 3-body observables 
are determined by a few universal functions of the angle $\xi$.
In particular, the binding energies of the Efimov states are solutions to 
a transcendental equation involving a single universal function of 
$\xi$ \cite{Efi71}. In this paper, we calculate this universal 
function for the case of 3 identical bosons
and apply Efimov's equation to the $^4$He trimer.
We also extend Efimov theory to atoms with deep 2-body bound states.

The existence of Efimov states can easily be understood in terms of the
equation for the radial wave function $f(R)$ in the adiabatic 
hyperspherical representation of the 3-body problem \cite{Jen93,Nie01}.
The hyperspherical radius for 3 identical atoms with coordinates
$\vec{r}_1$, $\vec{r}_2$, and $\vec{r}_3$ is $R^2=(r_{12}^2+r_{13}^2+
r_{23}^2)/3$, where $r_{ij}=|\vec{r}_i -\vec{r}_j|$.
If $|a| \gg l$, the radial equation for 3 atoms
with total angular momentum zero reduces in the region $l \ll R
\ll |a|$ to
\beq
-{\hbar^2 \over 2m} \left[ {\partial^2 \over \partial R^2}
+ {s_0^2 + 1/4 \over R^2} \right] f(R) = E f (R),
\label{radial}
\eeq
where $s_0 \approx 1.00624$.
This looks like the Schr{\"o}dinger equation for a particle in a 
one-dimensional, scale-invariant $1/R^2$ potential. 
If we impose a boundary condition on $f(R)$ 
at short-distances of order $l$, the radial equation
(\ref{radial}) has solutions at discrete negative values of the 
eigenvalue $E=-B_3$, with $B_3$
ranging from order $\hbar^2/m l^2$ to order 
$\hbar^2/ma^2$. The corresponding eigenstates are called Efimov states.
As $|a|\to \infty$, their spectrum approaches the simple
law $B_3^{(n)} \sim 515^{n}\hbar^2/ma^2$.

Efimov's constraints can be derived by constructing a solution to
Eq.~(\ref{radial}) that is valid in the region $l \ll R \ll |a|$. 
In the case of a bound state with energy 
$E=-B_3$, the radial variable is $H^2=mB_3/\hbar^2 +1/a^2$ and
the angular variable $\xi$ is
\bqa
\xi  &=& -\arctan (a\sqrt {mB_3}/ \hbar)
-\pi \theta(-a) \,,
\label{Hxi-def}
\eqa
where $\theta(x)$ is the unit step function.
Since we are interested in low energies $|E| \sim \hbar^2/ma^2$,
the energy eigenvalue in (\ref{radial}) can be neglected. The 
most general solution therefore has the form \cite{Efi71}
\beq
f (R) = \sqrt{H R} \left[A e^{is_0 \ln (H R)} + B e^{-is_0 \ln (H R)}
\right],
\label{f-general}
\eeq
which is the sum of outgoing and incoming hyperspherical waves. 
The dimensionless coefficients $A$ and $B$ can depend on $\xi$. 
At shorter distances $R \sim l$ and longer distances $R\sim |a|$,
the wavefunction becomes very complicated. 
Fortunately, we can avoid solving for it by using
simple considerations based on unitarity \cite{Efi71}.

We first consider the short-distance region. 
Efimov assumed implicitly that there are no deep 2-body
bound states with binding energies $B \gg \hbar^2/m a^2$. Thus the
2-body potential supports no bound states at all if $a<0$ and only the
dimer with binding energy $B_2=\hbar^2/ma^2$ if $a>0$. 
We will address the complication of deep 2-body bound states later. 
The probability in the incoming
wave must then be totally reflected by the potential at short distances,
so we can set $B=Ae^{i\theta}$.
The phase $\theta$ can be specified by giving the
logarithmic derivative $R_0 f' (R_0)/ f (R_0)$ at any
point $l \ll R_0 \ll |a|$. The resulting expression for $\theta$
has a simple dependence on $H$:
\bqa
\theta/2 &=&s_0 \ln (H/ c\Lambda_*)\,.
\label{theta-star}
\eqa
The denominator $c\Lambda_*$ is a complicated function of
$R_0$ and $R_0 f'(R_0)/f(R_0)$. It differs by an unknown 
constant $c$ from the 3-body parameter $\Lambda_*$ introduced in 
Ref.~\cite{BHK99}.

We next consider large distances $R\sim |a|$. In general, 
an outgoing hyperspherical wave incident on the $R\sim |a|$ region
can either be reflected or else transmitted to $R\to \infty$
as a 3-atom or atom-dimer scattering state.
The reflection and transmission amplitudes are described
by a dimensionless unitary $3\times 3$ matrix that is a 
function of $\xi$ only.
For $-\pi < \xi < -\pi/4$, scattering states with $R\to\infty$ are
kinematically not allowed. The probability is therefore totally reflected,
so we must have $B=Ae^{i\Delta(\xi)}$, where the phase
$\Delta$ depends on the angle $\xi$. Compatibility with the
constraint from short distances requires
$\theta = \Delta (\xi)  \; {\rm mod} \; 2\pi$.
Using Eq.~(\ref{theta-star}) for $\theta$ and inserting the
expression for $H$, we obtain Efimov's equation \cite{Efi71}
\beq
B_3  + {\hbar^2 \over ma^2} = \frac{\hbar^2 \Lambda_*^2}{m}\,
e^{2 \pi n/ s_0} \exp \left[\Delta \left( \xi \right)/s_0
\right]\,,
\label{B3-Efimov}
\eeq
where $\xi$ is given by (\ref{Hxi-def}) and the 
constant $c$ was absorbed into $\Delta(\xi)$. Note that we measure 
$B_3$ from the 3-atom threshold and $\Lambda_*$
is only defined up to factors of $\exp[\pi/s_0]$. Once the universal
function $\Delta(\xi)$ is known, the Efimov binding energies $B_3$
can be calculated by solving Eq.~(\ref{B3-Efimov}) for different 
integers $n$.
This equation has an exact discrete scaling symmetry: if there is an
Efimov state with binding energy $B_3$ for the parameters $a$
and $\Lambda_*$, then there is also an Efimov state with binding energy
$\lambda^2 B_3$ for the parameters $\lambda^{-1} a$ and $\Lambda_*$
if $\lambda=\exp[n'\pi/s_0]$ with $n'$ an integer. 

The universal function $\Delta(\xi)$ could be determined by
solving the 3-body equation for the Efimov
binding energies in various potentials whose scattering lengths
are so large that effective range corrections are negligible. It 
can be calculated more easily by using the effective field theory (EFT)
of Ref.~\cite{BHK99} in which the effective range can be set to zero.
In Ref.~\cite{BHK99}, the dependence of the binding energy
on $a$ and $\Lambda_*$ was calculated for the shallowest Efimov state
and $a>0$. In order to extract the universal function $\Delta(\xi)$,
we have calculated the binding energies of the three lowest
Efimov states for both signs of $a$. 
\begin{figure}[tb]
\begin{center}
\centerline{\includegraphics[width=13cm,angle=0,clip=true]{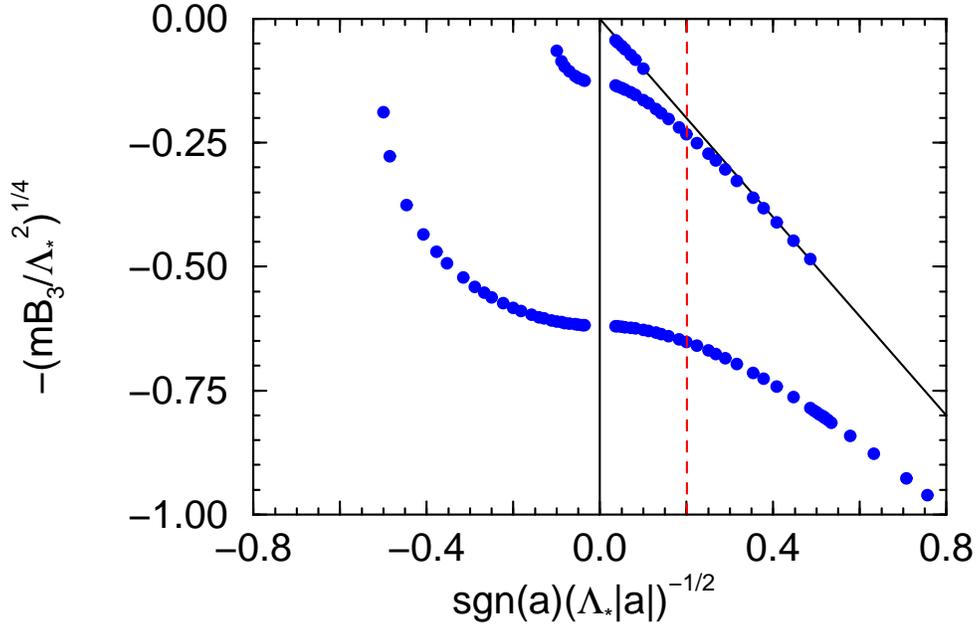}}
\end{center}
\vspace*{-25pt}
\caption{The energy variable $-(mB_3/\hbar^2\Lambda_*^2)^{1/4}$ for three 
 shallow Efimov states as a function of 
 ${\rm sgn }(a)(\Lambda_* |a|)^{-1/2}$.}
\label{fig:B3n}
\end{figure}
In Fig.~\ref{fig:B3n},
we plot $-(mB_3/\hbar^2\Lambda_*^2)^{1/4}$ as a function of 
${\rm sgn }(a)(\Lambda_* |a|)^{-1/2}$ for these three 
branches of Efimov states. 
The binding energies for deeper Efimov states and for shallower states
near $(\Lambda_* |a|)^{-1/2}=0$ can be obtained from the discrete 
scaling symmetry.
A given 2-body potential is characterized by values of $a$ and
$\Lambda_*$ and corresponds to a vertical line in Fig.~\ref{fig:B3n},
such as the dashed line shown.
The intersections of this line with the binding energy curves
correspond to the infinitely many Efimov states. Those states with 
$B_3 \simge \hbar^2/ml^2$ are unphysical.
For $B_3 \to \infty$, the angle $\xi$ goes to $-\pi/2$.
The ratio of the binding energies of successive Efimov states
therefore approaches $\exp[2\pi/s_0]\approx 515$.
However, for the shallowest Efimov states, this ratio
exhibits significant deviations from the asymptotic
value. If $a>0$, there is an Efimov state at the atom-dimer
threshold $B_3=\hbar^2/ma^2$ when $s_0 \ln(a\Lambda_*)=1.444\; {\rm mod}\;
\pi$. The sequence of binding energies $B_3$ in units of
$\hbar^2/ma^2$ is 1, 6.8, $1.4\times 10^{3}$,$\,...\,$. 
Consequently, the ratio of $B_3$ for the two shallowest Efimov states 
can range from 6.8 to 210. 
If $a<0$, there is an Efimov state at the 3-atom
threshold $B_3=0$ when $s_0 \ln(a\Lambda_*)=1.378\; {\rm mod}\; \pi$.
The sequence of binding energies $B_3$ in units of $\hbar^2/ma^2$ is 0,
$1.1 \times 10^3$, $6.0 \times 10^5$,$\,...\,$. Thus, 
the ratio of $B_3$ for the two shallowest Efimov states 
can range from $\infty$ to 550.

Using Eq.~(\ref{B3-Efimov}) with $n=0$, 
we have extracted the universal function
$\Delta(\xi)$ from the data for the middle branch in Fig.~\ref{fig:B3n}. 
The extracted values of $\Delta(\xi)$ are given in Table~\ref{tab0}.
\begin{table}[htb]
\begin{tabular}{c|c||c|c||c|c}
$\xi$ & $\Delta(\xi)$ & $\xi$ & $\Delta(\xi)$ & $\xi$ & $\Delta(\xi)$\\
\hline\hline
-0.785 &   -2.214 & -0.965 &   -5.712 & -1.482 &   -8.009\\
-0.787 &   -2.539 & -1.019 &   -6.123 & -1.502 &   -8.059\\
-0.791 &   -2.897 & -1.065 &   -6.415 & -1.651 &   -8.373\\
-0.797 &   -3.194 & -1.104 &   -6.634 & -1.681 &   -8.427\\
-0.804 &   -3.448 & -1.166 &   -6.943 & -1.745 &   -8.534\\
-0.820 &   -3.864 & -1.214 &   -7.151 & -1.817 &   -8.641\\
-0.836 &   -4.196 & -1.296 &   -7.461 & -1.988 &   -8.843\\
-0.852 &   -4.469 & -1.347 &   -7.632 & -2.197 &   -9.009\\
-0.868 &   -4.701 & -1.408 &   -7.814 & -2.395 &   -9.095\\
-0.899 &   -5.076 & -1.443 &   -7.910 & -2.751 &   -9.110\\
-0.933 &   -5.434 
\end{tabular}
\caption{The values of the universal function $\Delta(\xi)$.}
\label{tab0}
\end{table}
An analytic expression for $\Delta(\xi)$ is not known, but
we can obtain parametrizations in various regions of
$\xi$ by fitting the data: 
\bqa
\label{expol1}
&&\mqo2\mqo2\mqo2\xi \in {\textstyle [-{3\pi \over 8},-{\pi \over 4}]:}\,
\Delta=3.10x^2 -9.63x -2.18 ,\;\\[2pt]
\label{expol2}
&&\mqo2\mqo2\mqo2\xi \in {\textstyle [-{5\pi \over 8},-{3\pi \over 8}]:}\,
\Delta=1.17y^3+1.97y^2+2.12y-8.22,\quad\\[2pt]
\label{expol3}
&&\mqo2\mqo2\mqo2\xi \in {\textstyle [-\pi,-{5\pi \over 8}]:}\,
\Delta=0.25z^2+0.28z-9.11,\;
\eqa
where $x=(-\pi/4-\xi)^{1/2}$, $y=\pi/2+\xi$, and $z=(\pi+\xi)^2
\exp[-1/(\pi+\xi)^2]$.
These parametrizations deviate from the numerical results 
in Table \ref{tab0} by less than 
0.013. The discontinuity at $\xi=-{3\pi \over 8}$ and $\xi= -{5\pi \over 8}$
is less than 0.016.
This accuracy is sufficient for most practical calculations using
Eq.~(\ref{B3-Efimov}). 

\begin{figure}[tb]
\begin{center}
\centerline{\includegraphics[width=13cm,angle=0,clip=true]{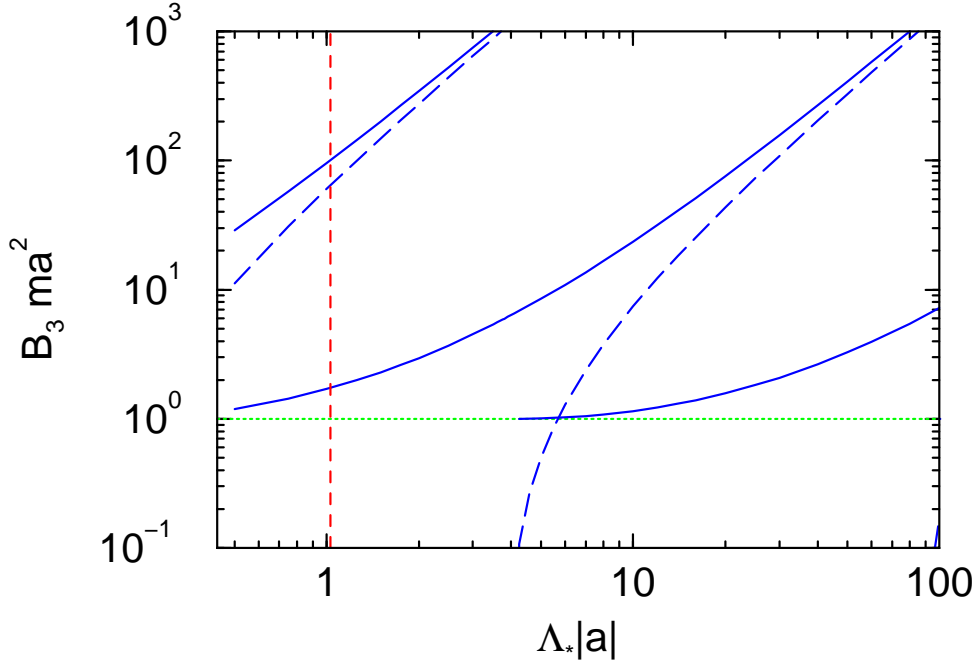}}
\end{center}
\vspace*{-25pt}
\caption{The Efimov binding energies $B_3 m a^2/\hbar^2$ 
as a function of $|a|\Lambda_*$ for $a>0$ (solid lines) and
$a<0$ (dashed lines). Vertical dashed line gives $|a|\Lambda_*$
for LM2M2/TTY potentials. Horizontal dotted line is the atom-dimer
threshold ($a>0$).}
\label{fig:efiB3}
\end{figure}

Universality can be exploited to greatly reduce the calculational
effort required to predict 3-body observables for atoms with
large $a$. The observables can be calculated once
and for all as functions of $a$ and $\Lambda_*$ either by using EFT or 
by solving the Schr{\"o}dinger or Faddeev equations with
various methods.
The binding energies obtained by solving Efimov's equation
(\ref{B3-Efimov}) are shown in Fig.~\ref{fig:efiB3}.
Simple expressions can be given for other observables,
such as the $S$-wave atom-dimer scattering length:
\beq
a_{12}=a\left( 1.46 - 2.15 \tan[s_0 \ln(a\Lambda_*)+ 0.09 ]\right)\,,
\label{a12-para} 
\eeq
as well as the phase shifts and the rate constant for 3-body 
recombination at threshold \cite{BrH02}.
Given $a$ and a measured or calculated value of $B_3$ for one Efimov state
as input, one can read off $\Lambda_*$ from Fig.~\ref{fig:efiB3}.
Predictions for other 3-body observables, such as the atom-dimer
scattering length in (\ref{a12-para}), are then immediate.

One of the most promising systems for observing Efimov states is
$^4$He atoms. The $^4$He trimer has been observed \cite{STo96}, but
no quantitative experimental information about its binding energy
is available to date. The binding energy has been calculated accurately 
for various model potentials. They indicate that there are two trimers,
a ground state with binding energy $B_3^{(0)}$ and an excited state
with binding energy $B_3^{(1)}$. The most accurate calculations have been
obtained by solving the Faddeev equations in
the hyperspherical representation \cite{NFJ98},
in configuration space \cite{RoY00},
and with hard-core boundary conditions \cite{MSSK01}.
These methods all give consistent results. The results of 
Ref.~\cite{MSSK01} for $B_2$, $B_3^{(0)}$, $B_3^{(1)}$, and the 
atom-dimer scattering length $a_{12}$ for four different
model potentials are given in Table~\ref{tab1}.
\begin{table}[htb]
\begin{tabular}{c||c|c|c|c||c|c|c}
Potential & $B_2$ & $B_3^{(0)}$ & $B_3^{(1)}$ & $a_{12}$ & 
$a_B \Lambda_*$ & $B_3^{(0)}$ & $a_{12}$ \\
\hline\hline
HFDHE2 & 0.830 & 116.7 & 1.67 & $-$    & 1.258 & 118.5 & 87.9 \\
HFD-B  & 1.685 & 132.5 & 2.74 & 135(5) & 0.922 & 137.5 & 120.2 \\
LM2M2  & 1.303 & 125.9 & 2.28 & 131(5) & 1.033 & 130.3 & 113.1 \\
TTY    & 1.310 & 125.8 & 2.28 & 131(5) & 1.025 & 129.1 & 114.5
\end{tabular}
\caption{The values of $B_2$, $B_3^{(0)}$, $B_3^{(1)}$, and $a_{12}$ 
for four model potentials from Ref.~\protect\cite{MSSK01},
the value of $a_B \Lambda_*$ determined from $B_3^{(1)}$, 
and the predictions for $B_3^{(0)}$ from Eq.~(\ref{B3-Efimov})
and $a_{12}$ from Eq.~(\ref{a12-para}). All energies (lengths) are 
given in mK (\AA). ($\hbar^2/m =12.12$ K$\,$\AA$^2$ for $^4$He).
}
\label{tab1}
\end{table}
The discrepancies between $B_2$ and the large-$a$ prediction $\hbar^2/ma^2$ 
are about 6-8\% \cite{MSSK01}. They can be
attributed to effective range corrections and provide estimates of
the error associated with the large-$a$ approximation.

We proceed to illustrate the power of universality by
applying it to  the $^4$He trimer. 
For the scattering length $a$, we take the value $a_B\equiv\hbar/\sqrt{mB_2}$
obtained from the calculated dimer binding energy.
We determine $\Lambda_*$ by demanding that $B_3^{(1)}$ satisfy
(\ref{B3-Efimov}) with $n=1$. Solving Eq.~(\ref{B3-Efimov}) with $n=2$,
we obtain the predictions for $B_3^{(0)}$ in the
second-to-last column of Table~\ref{tab1}. The predictions 
are only 1-4\% higher than the calculated values, which is 
within the expected error for the large-$a$ approximation.
This demonstrates that the ground state of the 
$^4$He trimer can be described by Efimov's equation (\ref{B3-Efimov}).
If we use the calculated values 
of $a$ as input instead of $B_2$, the predicted values of
$B_3^{(0)}$ are larger than the calculated values by 11-21\%.

We can use the value of $a_B \Lambda_*$ determined from the
excited state of the trimer to predict the atom-dimer scattering
length $a_{12}$ and compare with the calculated values in the fourth
column of Table~\ref{tab1}. 
The predictions for the four model potentials
are given in the last column of Table~\ref{tab1}. 
They are smaller than the calculated values by about 13\%. If the
calculated value of $a$ is used as input they are smaller by about 28\%.
It should be possible to account for these differences quantitatively
by taking into account higher order corrections \cite{BrH02}.

Efimov implicitly assumed in his derivation of Eq.~(\ref{B3-Efimov}) that 
there were no deep 2-body bound states \cite{Efi71}. 
If such states are present, the Efimov states become
resonances that can decay into a deep 2-body
bound state and a recoiling atom. Thus their energies are given by
complex numbers $E=-B_3-i\Gamma_3/2$.
If a potential supports many 2-body bound states,
the direct calculation of the widths $\Gamma_3$ by solving the 
Schr{\"o}dinger equation is very difficult \cite{NiE00}. 
However, one can show that the cumulative effect of all deep 2-body 
bound states on low-energy 3-body observables can be taken into account by 
including one additional low-energy parameter $\eta$.
Three-body recombination into a deep 2-body bound state
with binding energy $B\sim \hbar^2/ml^2$ can only take place
if $R\sim l$. It is obvious
that the atoms that form the bound state must approach to within a
distance of order $l$, since the size of the bound state is of
order $l$. However, the third atom must also approach the
pair to within a distance of order $l$, because it must recoil
with momentum $\sqrt{4mB /3}\sim \hbar/l$, and the  
necessary momentum kick can be delivered only if $R\sim l$.
The atoms can approach
such short distances only by following the lowest continuum adiabatic 
hyperspherical potential, because it is attractive in the region 
$l \ll R \ll |a|$, while all other potentials are repulsive.
Thus all pathways to final states including a deep 2-body bound 
state must flow through this lowest continuum adiabatic
potential. Note the lowest continuum state can either be a 3-atom state
or an  atom-dimer state. See, e.g., Fig.~5 in Ref.~\cite{Nie01} for an 
explicit calculation of these potentials.
The cumulative effects of 3-body recombination into deep
2-body bound states can therefore be described by the reflection probability
$e^{-\eta}$ for hyperspherical waves entering the region $R\sim l$
of this potential.
Up to corrections suppressed by $l/|a|$, the low-energy 3-body
observables are all determined by $a$, $\Lambda_*$, and $\eta$.  

We proceed to generalize Efimov's equation (\ref{B3-Efimov})
to the case in which there are deep 2-body
bound states. Again, we combine the analytic solution (\ref{f-general})
to the radial equation in the region $l \ll R \ll |a|$ with simple 
probability considerations at $R\sim |a|$ and $R\sim l$. 
Since the existence of deep 2-body bound states plays no role
in the unitarity constraint at $R\sim |a|$, we still
have $B=Ae^{i\Delta(\xi)}$ with the same universal function $\Delta(\xi)$
given in (\ref{expol1})-(\ref{expol3}). In the unitarity constraint
at $R\sim l$, we need to take into account that
only a fraction of the probability that flows to
short distances is reflected back to long distances
through the lowest adiabatic hyperspherical potential associated
with the 3-atom or atom-dimer continuum. Denoting the
reflection probability by $e^{-\eta}$, the coefficient $B$ in
(\ref{f-general}) can be written $B=Ae^{i\theta+\eta/2}$.
The compatibility condition $\theta-i\eta/2 = \Delta(\xi)\;
{\rm mod} \; 2\pi$ then becomes
\beq
B_3+{i\over 2}\Gamma_3 + {\hbar^2 \over ma^2} 
= \frac{\hbar^2 \Lambda_*^2}{m}\, 
e^{2\pi n/s_0} e^{(\Delta(\xi)+i\eta/2)/s_0},
\label{B3-Efimov-width}
\eeq
where $\xi$ is defined by (\ref{Hxi-def}) with $B_3 \to B_3 +i \Gamma_3/2$.
To solve this equation, we need the analytic continuation of 
$\Delta(\xi)$ to complex values of $\xi$.
The parametrization (\ref{expol1})-(\ref{expol3}) for $\Delta(\xi)$
should be accurate for complex $\xi$ with sufficiently small
imaginary parts, except perhaps near $\xi=-\pi$ where
it has an essential singularity.
If $B_3$ and $\Gamma_3$ for one Efimov state
are known, they can be used to determine $\Lambda_*$ and $\eta$.
The remaining spectrum of Efimov states and their widths can then be 
calculated by solving Eq.~(\ref{B3-Efimov-width}).
For infinitesimal $\eta$, the widths approach
$\Gamma_3 \to (\eta/s_0) ( B_3 +\hbar^2/ma^2 )\,$.
The widths of the deeper Efimov states increase geometrically just like the
binding energies, as has been observed in numerical calculations 
\cite{NiE00}. If $\eta$ is so large that $B_3 \sim \Gamma_3$,
the Efimov states cease to exist in any meaningful sense.

We have calculated the universal function $\Delta(\xi)$ that appears in 
Efimov's equation (\ref{B3-Efimov}) for the binding energies $B_3$ of
Efimov states. This equation can  be used as an operational definition 
of the 3-body parameter $\Lambda_*$ introduced in Ref.~\cite{BHK99}.
If $B_3$ for one Efimov state is known, it can be used to
determine $\Lambda_*$, and then  
universality predicts all low-energy 3-body observables as functions of
$a$ and $\Lambda_*$.
In Eq.~(\ref{B3-Efimov-width}), we have generalized Efimov's 
equation to permit deep 2-body bound states. The generalization
involves an additional inelasticity parameter $\eta$,
but the spectrum is determined by the same 
universal function $\Delta(\xi)$.

This research was supported by DOE Grant No. DE-FG02-91-ER4069 and
NSF Grant No. PHY-0098645.

\end{document}